\documentclass[10pt,conference,final]{IEEEtran}

\newcommand\copyrighttext{%
  \footnotesize \textcopyright 2021 IEEE. Personal use of this material is permitted. Permission from IEEE must be obtained for all other uses, in any current or future media, including reprinting/republishing this material for advertising or promotional purposes, creating new collective works, for resale or redistribution to servers or lists, or reuse of any copyrighted component of this work in other works.
  DOI: TBD}

\newcommand\copyrightnotice{%
\begin{tikzpicture}[remember picture,overlay]
\node[anchor=south,yshift=10pt] at (current page.south) {\fbox{\parbox{\dimexpr\textwidth-\fboxsep-\fboxrule\relax}{\copyrighttext}}};
\end{tikzpicture}%
}

\usepackage[utf8]{inputenc}
\usepackage[T1]{fontenc}
\usepackage{microtype}

\usepackage{amssymb}
\usepackage{amsmath}

\usepackage{graphicx}
\graphicspath{{fig/}}

\usepackage[table]{xcolor}
\usepackage{multicol}
\usepackage{multirow}
\usepackage{tabularx}
\usepackage{verbatim}
\usepackage{pgfplots}

\newcolumntype{C}{>{\centering\arraybackslash}X}
\newcolumntype{s}{>{\centering}m{0.1\columnwidth}}

\usepackage{listings}
\usepackage{syntax}
\usepackage{algorithm}
\usepackage[noend]{algpseudocode}

\usepackage{subcaption}
\usepackage{mdwlist}
\usepackage{tablefootnote}

\usepackage{todonotes}

\usepackage[all]{nowidow}
\usepackage{balance}

\usepackage{url}

\lstdefinestyle{basic}{
  extendedchars     = true,
  inputencoding     = utf8,
  basicstyle        = {\ttfamily \small},
  keywordstyle      = {\rmfamily \bfseries},
  commentstyle      = {\rmfamily \itshape},
  tabsize           = 2,
  flexiblecolumns   = false,
  frame             = single,
  showstringspaces  = false,
  breaklines        = true,
}

\lstdefinelanguage{Kotlin}{
  keywords = {
    package, as?, typealias,
    this, super, val, var,
    fun, for, null, true,
    false, throw,
    return, break, continue, object,
    if, try, else, while,
    do, when, class, interface,
    enum, companion, override, public,
    private, get, set, import,
    abstract, vararg, expect, actual,
    where, suspend, data, internal,
    dynamic, final, by
  },
  keywordstyle = {\bfseries},
  ndkeywords = {
    @Deprecated, @JvmName, @JvmStatic, @JvmOverloads,
    @JvmField, @JvmSynthetic, Iterable, Int,
    Long, Integer, Short, Byte,
    Float, Double, String, Runnable,
    Array
  },
  ndkeywordstyle = {\bfseries},
  emph = {
    println, return@, forEach, map,
    mapNotNull, first, filter, firstOrNull,
    lazy, delegate
  },
  emphstyle       = {},
  identifierstyle = \color{black},
  sensitive       = true,
  commentstyle    = {\color{gray}\ttfamily},
  comment         = [l]{//},
  morecomment     = [s]{/*}{*/},
  stringstyle     = {\ttfamily},
  morestring      = [b]",
  morestring      = [s]{"""*}{*"""},
}

\lstdefinestyle{src-cpp}{
  style    = basic,
  language = C++,
}

\lstdefinestyle{pascal}{
	style    = basic,
	language = Pascal,
}

\lstdefinestyle{msg}{
    style    = basic,
	language = Kotlin,
}

\begin{document}

\title{Type-Centric Kotlin Compiler Fuzzing: Preserving Test Program Correctness by Preserving Types}

\author{
\IEEEauthorblockN{Daniil Stepanov\IEEEauthorrefmark{1}, Marat Akhin\IEEEauthorrefmark{2}, and Mikhail Belyaev\IEEEauthorrefmark{3}}
\IEEEauthorblockA{\textit{Peter the Great St.\,Petersburg Polytechnic University}\\
\textit{Jetbrains Research}\\
Saint Petersburg, Russia\\
\IEEEauthorrefmark{1}\url{stepanov@kspt.icc.spbstu.ru}\\
\IEEEauthorrefmark{2}\url{akhin@kspt.icc.spbstu.ru}\\
\IEEEauthorrefmark{3}\url{belyaev@kspt.icc.spbstu.ru}
}
}

\maketitle
\copyrightnotice

\begin{abstract}
Kotlin is a relatively new programming language from JetBrains: its development started in 2010 with release 1.0 done in early 2016.
The Kotlin compiler, while slowly and steadily becoming more and more mature, still crashes from time to time on the more tricky input programs, not least because of the complexity of its features and their interactions.
This makes it a great target for fuzzing, even the basic forms of which can find a significant number of Kotlin compiler crashes.

There is a problem with fuzzing, however, closely related to the cause of the crashes: generating a random, non-trivial and semantically valid Kotlin program is hard.
In this paper, we talk about \emph{type-centric compiler fuzzing} in the form of \emph{type-centric enumeration}, an approach inspired by skeletal program enumeration~\cite{zhang2017skeletal} and based on a combination of generative and mutation-based fuzzing, which solves this problem by focusing on program \emph{types}.
After creating the skeleton program, we fill the typed holes with fragments of suitable type, created via generation and enhanced by semantic-aware mutation.

We implemented this approach in our Kotlin compiler fuzzing framework called Backend Bug Finder~(BBF) and did an extensive evaluation, not only testing the real-world feasibility of our approach, but also comparing it to other compiler fuzzing techniques.
The results show our approach to be significantly better compared to other fuzzing approaches at generating semantically valid Kotlin programs, while creating more interesting crash-inducing inputs at the same time.
We managed to find more than 50 previously unknown compiler crashes, of which 18 were considered important after their triage by the compiler team.
\end{abstract}


\section{Introduction}
\label{sec:introduction}

Compilers are one of the most important tools in software development, and their correctness is therefore a very important topic.
However, just as most other software systems, compilers have bugs of different kinds, starting from performance degradations on specific program inputs and ending with miscompilations, possibly leading to severe problems in user software.
Fuzzing~\cite{chen2013taming, marcozzi2019fuzzingmatters} has emerged in recent years as one of the most prominent methods of checking compiler correctness.

One may believe compiler fuzzing to be an easy process: how hard can it be to generate a random program, compile it and check the results?
Unfortunately, finding bug-inducing non-trivial programs is hard~\cite{chen2020survey}, despite all the recent advances in fuzzing.
Generating ``interesting'' programs, which uncover bugs relevant to the compiler developers, is even more difficult.

In our experience from working with the Kotlin language%
\footnote{\url{https://kotlinlang.org/}}
compiler team, actionable bugs are usually bugs closest to being real-world programs, i.e., syntactically and semantically valid, feasible to be written by an actual developer, and inducing a crash.
Generational fuzzers~\cite{kreutzer2020language, yang2011finding}, while better at making semantically valid programs, usually fail to create human-looking code.
Mutation-based approaches~\cite{veggalam2016ifuzzer, holler2012fuzzing}, on the other hand, are good at mimicking real developers, but often generate semantically incorrect programs.

Skeletal program enumeration~\cite{zhang2017skeletal} is a very promising mutation-based approach, targeted at generating more semantically correct programs.
It is built on the following idea: consider a given program with its variables replaced by placeholders, creating a \emph{skeleton}.
By filling the placeholders with original variables in different permutations, we can enumerate over possible programs to efficiently fuzz the compiler.

This approach has some obvious problems, if we are working with a language having a complex type systems, as many permutations will be incorrect w.r.t. placeholder types.
This observation lead us to the idea of \emph{type-centric enumeration}: instead of considering placeholders as unfilled variables, we consider placeholders as unfilled \emph{types}.
To handle them, we first generate expressions of required type, similarly to existing generational fuzzing approaches.
Then, in the mutation phase, we fill the placeholders with the generated expressions, taking into consideration their types.

We have implemented our approach inside a tool for Kotlin compiler fuzzing named Backend Bug Finder%
\footnote{\url{https://github.com/vorpal-research/bbf}}%
~(BBF) and performed two series of experimental evaluation.
For the first set of experiments, we ran type-centric enumeration for 2 weeks and found more than 50 previously unknown Kotlin compiler bugs, 18 of which were considered interesting after their triage in the Kotlin bug tracker.
To compare our approach against existing ones, in the second part of evaluation we ran it and a number of other fuzzing techniques for 12 hours.
The results show type-centric enumeration to outperform alternative approaches, if one is concerned with finding interesting bugs and miscompilations.
However, if you want to thoroughly test your compiler, you should use several complimentary approaches, to fuzz different parts of the compiler efficiently.

The rest of the paper is organized as follows. We describe our approach in section~\ref{sec:basics}.
In section~\ref{sec:implementation} we explain the most important details of our implementation and what practical challenges we encountered.
Evaluation, its results and their analysis are discussed in section~\ref{sec:evaluation}.
We talk about related work in section~\ref{sec:related-work}, make conclusions and briefly talk about plans for future work in section~\ref{sec:conclusion}.

\section{Type-centric fuzzing}
\label{sec:basics}

\begin{figure}[tb]
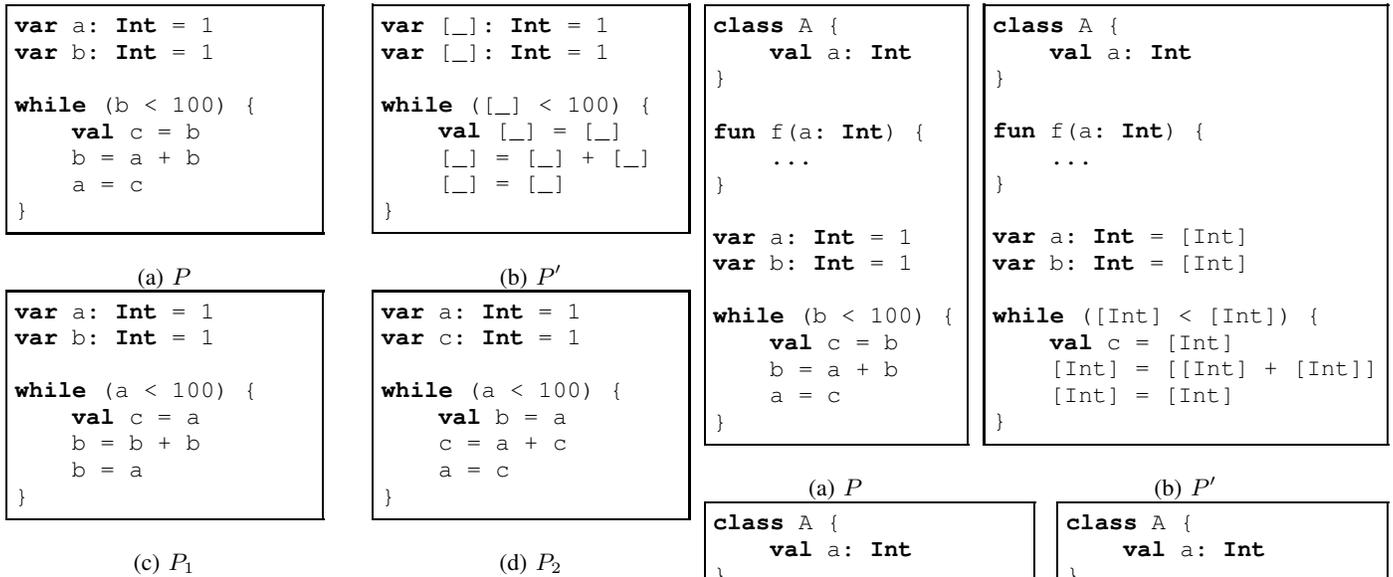

\begin{subfigure}[b]{0.45\linewidth}
\begin{lstlisting}[
  language=Kotlin,
  escapechar=|
]
var a: Int = 1
var b: Int = 1

while (b < 100) {
    val c = b
    b = a + b
    a = c
}
\end{lstlisting}
\caption{$P$}
\end{subfigure}%
\hfill%
\begin{subfigure}[b]{0.45\linewidth}
\begin{lstlisting}[
  language=Kotlin,
  escapechar=|
]
var [_]: Int = 1
var [_]: Int = 1

while ([_] < 100) {
    val [_] = [_]
    [_] = [_] + [_]
    [_] = [_]
}
\end{lstlisting}
\caption{$P'$}
\end{subfigure}
\begin{subfigure}[b]{0.45\linewidth}
\begin{lstlisting}[
  language=Kotlin,
  escapechar=|
]
var a: Int = 1
var b: Int = 1

while (a < 100) {
    val c = a
    b = b + b
    b = a
}
\end{lstlisting}
\caption{$P_1$}
\end{subfigure}%
\hfill%
\begin{subfigure}[b]{0.45\linewidth}
\begin{lstlisting}[
  language=Kotlin,
  escapechar=|
]
var a: Int = 1
var c: Int = 1

while (a < 100) {
    val b = a
    c = a + c
    a = c
}
\end{lstlisting}
\caption{$P_2$}
\end{subfigure}

\caption{Skeletal program enumeration example}
\label{fig:spe-example}
\end{figure}

\subsection{Skeletal program enumeration}
\label{sec:spe}

Before we discuss our approach in more detail, we need to take a look at skeletal program enumeration~(SPE)~\cite{zhang2017skeletal}, which is one of the main inspirations of our work.
In SPE, the program~$P$ is transformed into a syntactic \emph{skeleton}~$P'$, with placeholders~(or holes) located at places of using \emph{variables} in the original program.
These placeholders are later filled with variables available from the set of all variables in the program, thus \emph{enumerating} over the space of possible skeleton instances.
An example of how SPE works is shown in figure~\ref{fig:spe-example}.

This idea looks simple and powerful enough, however it does have some intrinsic problems and difficulties.
First, a fairly large number of the resulting programs are semantically equivalent and must be eliminated.
To achieve this, the authors introduce the notion of \emph{$\alpha$-equivalent} programs: programs equivalent up to the renaming of their variables%
\footnote{This notion is similar to $\alpha$-conversion from lambda calculus~\cite{church1985lambda}}.
Enumerating only $\alpha$-equivalent programs allows SPE to significantly improve its performance.
Second, when working with most programming languages, one must take into account variable scopes, as otherwise many skeleton instances will be semantically incorrect~(due to scope errors).
This problem is handled by limiting the variable enumeration for a scope to variables available in this scope.

After implementing their approach and evaluating it on GCC test suite, the authors found more than 200 bugs in GCC/Clang, which confirms the applicability of SPE to real-world programs.

\begin{figure}
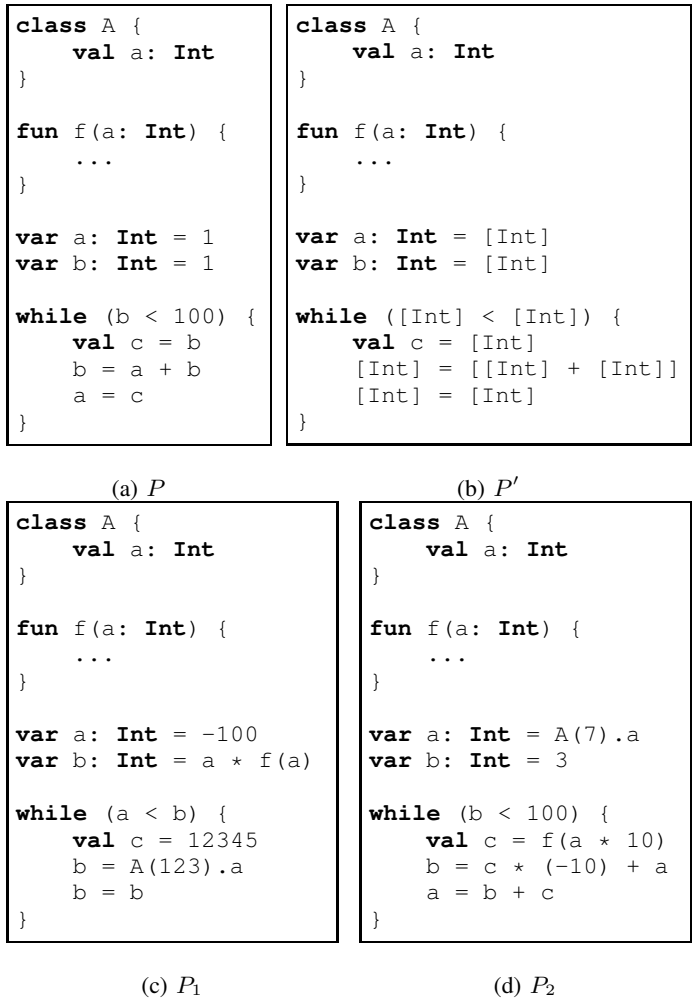

\begin{subfigure}[b]{0.37\linewidth}
\begin{lstlisting}[
  language=Kotlin,
  escapechar=|
]
class A {
    val a: Int
}

fun f(a: Int) {
    ...
}

var a: Int = 1
var b: Int = 1

while (b < 100) {
    val c = b
    b = a + b
    a = c
}
\end{lstlisting}
\caption{$P$}
\end{subfigure}%
\hfill%
\begin{subfigure}[b]{0.58\linewidth}
\begin{lstlisting}[
  language=Kotlin,
  escapechar=|
]
class A {
    val a: Int
}

fun f(a: Int) {
    ...
}

var a: Int = [|Int|]
var b: Int = [|Int|]

while ([|Int|] < [|Int|]) {
    val c = [|Int|]
    [|Int|] = [[|Int|] + [|Int|]]
    [|Int|] = [|Int|]
}
\end{lstlisting}
\caption{$P'$}
\end{subfigure}

\begin{subfigure}[b]{0.47\linewidth}
\begin{lstlisting}[
  language=Kotlin,
  escapechar=|
]
class A {
    val a: Int
}

fun f(a: Int) {
    ...
}

var a: Int = -100
var b: Int = a * f(a)

while (a < b) {
    val c = 12345
    b = A(123).a
    b = b
}
\end{lstlisting}
\caption{$P_1$}
\end{subfigure}%
\hfill%
\begin{subfigure}[b]{0.47\linewidth}
\begin{lstlisting}[
  language=Kotlin,
  escapechar=|
]
class A {
    val a: Int
}

fun f(a: Int) {
    ...
}

var a: Int = A(7).a
var b: Int = 3

while (b < 100) {
    val c = f(a * 10)
    b = c * (-10) + a
    a = b + c 
}
\end{lstlisting}
\caption{$P_2$}
\end{subfigure}

\caption{Type-centric enumeration example}
\label{fig:tce-example}
\end{figure}

\subsection{Type-centric enumeration}

Our approach (called type-centric enumeration or TCE) is based on SPE and uses the following intuition.
Instead of filling the skeleton placeholders with variables, we propose to fill them with \emph{types} in the form of typed generated expressions.
This allows us to sidestep the need for $\alpha$-equivalence, as most (if not all) skeleton instances will be semantically distinct, due to the variety of possible expressions.

Starting with a program~$P$, it is transformed into a typed skeleton~$P'$ with type placeholders~$p$ of type~$T$.
Additionally, we have sets of program variables~$V$ and of available callables~$C$~(such as functions, property accessors, methods, etc.).
The problem of creating a skeleton instance~$P_i$ is then reduced to filling type placeholders with expressions of suitable type, created with variables from~$V$ and callables from~$C$.

Consider the example from figure~\ref{fig:tce-example}.
Our skeleton~$P'$ has a number of \texttt{Int} placeholders, with $V = \{a, b, c\}$ and $C = \{\texttt{class A, accessor A::a, fun f}\}$.
Using these sets, for every \texttt{Int} placeholder we generate expressions of compatible type.
To do so, we use a combination of the following steps.
\begin{itemize}
    \item Take a variable $v$ available at the current scope from $V$;
    \item Generate a valid call from $C$;
    \item Create a random value of a supported primitive type~(booleans, integers, strings, etc.);
    \item Generate a valid call to the standard library.
\end{itemize}
For example, for $V = \{a\}$, $C = \{\texttt{fun f}\}$, support for primitive integers and standard library with integer operator~$+$, we could generate an expression \texttt{a + f(7) + 7} and use it to fill any $p$ of type~\texttt{Int}.
As the result of TCE, we get an possibly infinite number of semantically correct programs $P_i$, which can be used for compiler fuzzing.

TCE can be divided into 2 steps: typed expression generation~(aka generation phase) and type placeholder filling~(aka mutation phase).
Let us describe these steps in more details.

\begin{figure}[tb]
\begin{subfigure}[t]{\linewidth}
\begin{lstlisting}[
  language=Kotlin,
  escapechar=|
]
class A(val a: Int) {
    fun f(a: String): Int { ... }
}

interface B {
    val a: Int
}

class C(override val a: Int) : B

fun f(a: Int): Int

val a: Int = 1
\end{lstlisting}
\caption{Seed program for generation phase}
\label{fig:gen-example-1}
\end{subfigure}
\begin{subfigure}[t]{\linewidth}
\begin{lstlisting}[
  language=Kotlin,
  escapechar=|
]
A(1) -> |\textbf{A}|
A(1).a -> Int
A(1).f("") -> Int
C(1) -> |\textbf{B}|
C(1) -> |\textbf{C}|
C(1).a -> Int
f(1) -> Int
a -> Int
\end{lstlisting}
\caption{Generated expressions with their types}
\label{fig:gen-example-2}
\end{subfigure}
\caption{Generation phase example}
\label{fig:gen-example}
\end{figure}

\newcommand{\adaptTypeParameters}{\text{adaptTypeParams}}%
\newcommand{\args}{\mathit{args}}%
\newcommand{\callee}{\mathit{callee}}%
\newcommand{\call}{\mathit{call}}%
\newcommand{\ffile}{\mathit{file}}%
\newcommand{\findImplementation}{\text{findImplementation}}%
\newcommand{\generateCall}{\text{genCall}}%
\newcommand{\generateClassInstance}{\text{genClassInstance}}%
\newcommand{\generateConstructorCall}{\text{genConstructorCall}}%
\newcommand{\generateConstructor}{\text{genConstructor}}%
\newcommand{\generateInstance}{\text{genInstance}}%
\newcommand{\generateTypeParams}{\text{genTypeParams}}%
\newcommand{\generateValue}{\text{genValue}}%
\newcommand{\getCallables}{\text{getCallables}}%
\newcommand{\getInstanceCallables}{\text{getInstanceCallables}}%
\newcommand{\getRandomConstructor}{\text{getRandomConstructor}}%
\newcommand{\hasOpenConstructor}{\text{hasOpenConstructor}}%
\newcommand{\iCallee}{\mathit{iCallee}}%
\newcommand{\impl}{\mathit{impl}}%
\newcommand{\instance}{\mathit{instance}}%
\newcommand{\is}{\ \mathit{is}\ }%
\newcommand{\klass}{\mathit{class}}%
\newcommand{\parameterize}{\text{parameterize}}%
\newcommand{\randomCtor}{\mathit{randomCtor}}%
\newcommand{\res}{\mathit{res}}%
\newcommand{\typeParams}{\mathit{typeParams}}%
\newcommand{\usages}{\mathit{usages}}%

\algdef{SE}[FUNCTION]{Function}{EndFunction}%
   [2]{\algorithmicfunction\ \textproc{#1}\ifthenelse{\equal{#2}{}}{}{(#2)}}%
   {\algorithmicend\ \algorithmicfunction}

\begin{figure}[tb]
\setlength{\leftskip}{0cm}
\textbf{INPUT:} $\ffile$ with a seed program $P$ \\
\textbf{OUTPUT:} list of generated expressions $c_0...c_N$ \\
\begin{algorithmic}[1]

\Function{generationPhase}{$\ffile$}
    \State $\res \leftarrow \left[ \, \right]$
    
    \For{$\callee \in \getCallables(\ffile)$}
        \If{$\callee \is Class$}
            \State $\instance \leftarrow \generateClassInstance(\callee, \ffile)$
            \For{$\iCallee \in \getCallables(\instance)$}
                \State $\res \leftarrow \res + \generateCall(\iCallee, \ffile)$
            \EndFor
        \Else
            \State $\res \leftarrow \res + \generateCall(\callee, \ffile)$
        \EndIf
    \EndFor
    \State \Return $\res$
\EndFunction

\\
\Function{genClassInstance}{$\klass, \ffile$}
    \State $\typeParams \leftarrow \generateTypeParams(\klass, \ffile)$
    \State $\klass \leftarrow \parameterize(\klass, \typeParams)$
    \If{$ \neg \hasOpenConstructor(\klass)$}
        \State $\impl \leftarrow \findImplementation(\klass)$
        \If{$ \impl \neq null $}
            \State $\impl \leftarrow \adaptTypeParameters($
            \State $\qquad \impl, \klass, \typeParams)$
            \State \Return $\generateClassInstance(\impl, \ffile)$
        \Else
            \State \Return $null$
        \EndIf
    \EndIf
    \State $\randomCtor \leftarrow \getRandomConstructor(\klass)$
    \State \Return $\generateConstructorCall(\randomCtor, \ffile)$
\EndFunction


\end{algorithmic}
\caption{Generation phase algorithm}
\label{alg:gen-algo}
\end{figure}

\subsection{Generation phase}
\label{sec:generation-phase}

As input, the generation phase accepts a seed program~$P$ with available callables~$C$.
From this program, we need to generate a set of typed expressions using the callables from~$C$.

An example of such seed program is shown in figure~\ref{fig:gen-example-1}.
\texttt{class A} has a field~\texttt{a} of type~\texttt{Int} and a function~\texttt{f} with return type~\texttt{Int}.
In order to be able to call function~\texttt{f} from~\texttt{class A}, we must first generate an instance of~\texttt{class A}, generate the arguments of function~\texttt{f}, and then generate the call itself, as in, for example, \texttt{A(1).f("")}.
For object-oriented languages with inheritance, we need to also support it; for our example, it means understanding~\texttt{class C} is a implementation of~\texttt{interface B}.

Possible generated expressions are shown in figure~\ref{fig:gen-example-2}.
Of course, for real programs, the callables are more complicated than in the presented example, and the process and the results of generation would be much more difficult.

The high-level overview of the generation phase algorithm is shown in figure~\ref{alg:gen-algo}.
The algorithm begins with extraction of all callables~$c_0...c_N$ from the program.
Then, we enumerate over the expressions creatable from the extracted callables.

For classes, we attempt to generate its instance via~$\generateClassInstance$ function.
If a class has type parameters, we first need to generate them.
This process must consider the following important aspects.
\begin{itemize}
    \item Type parameters may be upper and/or lower bounded, which should be correctly handed to preserve the semantic correctness of the generated class instance;
    \item We need to limit the depth of nested type parameters; for example, at each iteration we may decrease the probability of generating parameterized types as type parameters;
    \item Recursive type parameters, if generated, should be handled with additional care.
\end{itemize}
After the generation, we must populate the class with the type parameters.
For example, if we generated~\texttt{Int} type parameter for~\texttt{class A<T>(val a: T)}, the result is~\texttt{class A<Int>(val a: Int)}.

If the class we are processing cannot be instantiated directly~(i.e., it is an abstract class or an interface), we attempt to find a suitable implementation.
This step also needs to support working with the standard library.
For example, if we need to generate an instance of~\texttt{Iterable<Int>}, we could do it by finding an implementation from the standard library, e.g.,~\texttt{ArrayList<Int>}.
Afterwards we adapt the previously generated type parameters~(if present) to the discovered implementation, as the implementation may introduce additional bounds and/or type parameters.
If all steps succeeded, we select a random constructor for the class and generate a call to it by generating values for its arguments.

After we have created an instance, we enumerate over all its available callables and generate calls to them, similarly to how we generate constructor calls by generating argument values.
This step finishes the generation of expressions obtainable for a given class on a single TCE iteration.

Generation of expressions for top-level callables skips over the creation of class instance and goes straight to the creation of a specific call.

\newcommand{\anon}{\mathit{anon}}%
\newcommand{\anonymize}{\text{anonymize}}%
\newcommand{\eType}{\mathit{eType}}%
\newcommand{\gen}{\mathit{gen}}%
\newcommand{\genPhExpr}{\text{genPhExpr}}%
\newcommand{\genRandomValue}{\text{genRandomValue}}%
\newcommand{\genStdLib}{\text{genStdLib}}%
\newcommand{\getPlaceholders}{\text{getPlaceholders}}%
\newcommand{\getType}{\text{getType}}%
\newcommand{\merge}{\text{merge}}%
\newcommand{\replacePhWithExpr}{\text{replacePhWithExpr}}%
\newcommand{\seed}{\mathit{seed}}%
\newcommand{\type}{\mathit{type}}%

\begin{figure}[tb]
\setlength{\leftskip}{0cm}
\textbf{INPUT:} seed program for mutation phase $\seed$ \\
\textbf{INPUT:} seed program from generation phase $\gen$ \\
\textbf{INPUT:} generated expressions $exprs$ \\
\textbf{OUTPUT:} program will filled type placeholders \\
\begin{algorithmic}[1]

\Function{mutationPhase}{}
    \State $\anon \leftarrow \anonymize(\seed)$
    \State $\anon \leftarrow \merge(\anon, \gen)$
    \For{$ph \in \getPlaceholders(\anon)$}
        \State $e \leftarrow \genPhExpr(ph, exprs)$
        \State $\anon \leftarrow \replacePhWithExpr(anon, ph, e)$
    \EndFor
    \State \Return{$\anon$}
\EndFunction

\\
\Function{genPhExpr}{$ph, exprs$}
    \State $\type \leftarrow \getType(ph)$
    \State $r \leftarrow \left[ \, \right]$
    \State $r \leftarrow r + \genRandomValue(\type)$
    \State $r \leftarrow r + \genStdLib(\type)$
    \For{$e \in exprs$}
        \State $\eType \leftarrow \getType(ph)$
        \If{$compatible(\type, \eType)$}
            \State $r \leftarrow r + e$
        \EndIf
    \EndFor
    \State \Return{$random(r)$}
\EndFunction
\end{algorithmic}
\caption{Mutation phase algorithm}
\label{alg:mut-algo}
\end{figure}

\begin{figure}[tb]
\begin{subfigure}[t]{\linewidth}
\begin{lstlisting}[
  language=Kotlin,
  escapechar=|
]
val a: Int = 1

class A(val a: Int) {
    fun f(a: String): Int { ... }
}

fun f(a: Int): Int { ... }

// Generated expressions with their types:
//   A(1) -> A
//   A(1).a -> Int
//   A(1).f("") -> Int
//   f(1) -> Int
//   a -> Int

\end{lstlisting}
\caption{Seed program after generation phase}
\label{fig:mut-example-1}
\end{subfigure}

\begin{subfigure}[t]{\linewidth}
\begin{lstlisting}[
  language=Kotlin,
  escapechar=|
]
fun factorial(n: Int): Double {
    var result = 1.0
    for (i in 1..n) {
        result *= i
    }
    return result
}
\end{lstlisting}
\caption{Seed program for mutation phase}
\label{fig:mut-example-2}
\end{subfigure}

\begin{subfigure}[t]{\linewidth}
\begin{lstlisting}[
  language=Kotlin,
  escapechar=|
]
fun factorial(n: Int): Double {
    var result = [|Double|]
    for (i in [|IntRange|]) {
        [|Double|] *= [|Int|]
    }
    return [|Double|]
}
\end{lstlisting}
\caption{Typed skeleton}
\label{fig:mut-example-3}
\end{subfigure}

\begin{subfigure}[t]{\linewidth}
\begin{lstlisting}[
  language=Kotlin,
  escapechar=|
]
val a: Int = 1

class A(val a: Int) {
    fun f(a: String): Int { ... }
}

fun f(a: Int): Int { ... }

fun factorial(n: Int): Double {
    var result = f(1).toDouble()
    for (i in A(1).f("")..a) {
        result *= i
    }
    return result
}
\end{lstlisting}
\caption{Typed skeleton filled with the generated expressions}
\label{fig:mut-example-4}
\end{subfigure}
\caption{Mutation phase example}
\label{fig:mut-example}
\end{figure}

\subsection{Mutation phase}
\label{sec:mutation-phase}

Assume the generation phase returned a set of expressions~$E$ as shown in figure~\ref{fig:mut-example-1}.
During the mutation phase, we use these generated expressions to fill the type placeholders in a second seed program, used as skeleton.
To transform the program to a typed skeleton, we perform type analysis, which calculates the types for selected type placeholders.
An example of a seed program and its typed skeleton are shown in figures~\ref{fig:mut-example-2} and~\ref{fig:mut-example-3}. 

Algorithm of the mutation phase is outlined in figure~\ref{alg:mut-algo}.
First, we anonymize the program names, to avoid possible name clashes.
Then, we merge the seed program from the generation phase to the seed program for the mutation phase, so that the generated expressions remain semantically valid.

After these preliminary steps are done, we begin filling out the placeholders.
For each placeholder of type~$T$ we generate an expression of a compatible type.
This can be done in one of the following ways.
\begin{itemize}
    \item Select a type-compatible expression from~$E$;
    \item Create a random value of a supported primitive type~(booleans, integers, strings, etc.);
    \item Generate a valid call to the standard library.
\end{itemize}

Let us consider the last item in more details.
If type~$T$ is somehow related to the standard library, we can generate an expression of suitable type as we do in the generation phase, but with callables including the relevant fragments from the standard library.
For example, if we need type~\texttt{List<Int>}, we could use a constructor of one of~\texttt{List} implementations, like~\texttt{ArrayList}, or any standard library function with compatible return type, like~\texttt{listOf<Int>()}.
Another way to use standard library is to increase variety in how we fill the placeholders.
If we know, for example, about the operator~\texttt{+} for~\texttt{List}, we can generate such expression as~\texttt{listOf<Int>(1) + listOf<Int>(1, 2, 3)}.

To increase variety even further, we may run the mutation phase not once, but several times.
To ensure termination, on each subsequent iteration we reduce the percentage of expressions in the program replaced by placeholders.
Additionally, after each mutation phase, we check if the semantic correctness of the program is preserved and roll back in case of an error.

\subsection{TCE problems}
\label{sec:tce-problems}

There are several important problems in TCE which must be solved for its successful application.
The first problem we briefly mentioned in section~\ref{sec:generation-phase} is that the generation process could potentially never terminate when generating type parameters.
The cause of the problem is that a type parameter may itself require a number of other type parameters.
In some cases, this process may not terminate, i.e., creating an infinite type parameter chain \texttt{class A<A<A...>\nobreak\hspace{.0001 em}>\nobreak\hspace{.0001 em}}.
To solve this problem, TCE reduces the probability of generating a type parameter with nested type parameters on each step.

The next problem we outlined in section~\ref{sec:mutation-phase} is, if one decides to use TCE in an iterative fashion and apply the mutation phase several times, the run time could increase non-linearly.
For example, if we fill a placeholder of type~\texttt{Int} with an expression~\texttt{listOf<Int>(1, 2, 3).get(0)}, on the subsequent mutation phase we have~4 possible new placeholders instead of~1, thus creating an exponential increase in run time.
As a workaround, we decided to use the following heuristic: on each iteration of the mutation phase we ensure the total number of placeholders is decreasing by a given ratio.
As a measure of last resort, we can also prematurely terminate the mutation process, if it takes too long.

The last important problem concerns generation of long chains of callables, both from the seed program and from the standard library.
Similarly to the type parameter problem, we could create a non-terminating sequence of calls, e.g., to create an expression of type~\texttt{Int}, we can generate it from~\texttt{List<Int>} by getting an element by index of type~\texttt{Int}, which we also need to generate, creating a possible infinite generation sequence.
To solve this problem, we limit the nested depth by an arbitrary constant, after which we do not consider callables from the standard library or the seed program for the generation process.

\begin{figure*}
\begin{subfigure}[b]{0.31\linewidth}
\begin{lstlisting}[
  language=Kotlin,
  escapechar=|
]
fun box() {
    val l = ArrayList<Long>()
    l.add(5.inv())
}
\end{lstlisting}
\caption{IR backend crash~(KT-42092)}
\label{fig:res-example-1}
\end{subfigure}%
\hfill
\begin{subfigure}[b]{0.35\linewidth}
\begin{lstlisting}[
  language=Kotlin,
  escapechar=|
]
fun box() {
    when ("abcd".sumOf { 1L }) {
        in 0..1 -> "A"
        else -> "B"
    }
}
\end{lstlisting}
\caption{IR backend crash~(KT-42054)}
\label{fig:res-example-2}
\end{subfigure}%
\hfill
\begin{subfigure}[b]{0.25\linewidth}
\begin{lstlisting}[
  language=Kotlin,
  escapechar=|
]
fun box() = 
    fun() = 
        ::intArrayOf
\end{lstlisting}
\caption{``Uninteresting'' bug example}
\label{fig:res-example-3}
\end{subfigure}
\caption{Examples of bugs found by TCE}
\label{fig:res-example}
\end{figure*}
\section{Implementation}
\label{sec:implementation}

We have implemented our approach inside a tool for Kotlin compiler fuzzing named Backend Bug Finder~(BBF).
In this section we discuss the most important details of the implementation, some of which are Kotlin-specific, and the others are completely language-agnostic.

\subsection{Working with the compiler}

To implement TCE, we need deep support from the language compiler.
First, as we need information about types, we should be able to utilize the type analysis subsystem of the compiler%
\footnote{This means TCE is best suited for statically typed languages, however, supporting dynamically typed languages is also possible with additional effort.}.
Second, as both the generation and the mutation phases actively change the code, we need access to some source code representation, e.g., abstract syntax tree~(AST).
Third, to access the standard library, it should be available in some shape or form through the compiler.

For these reasons, we decided to build BBF on top of Kotlin compiler.
Both the user code and the standard library are represented as Program Structure Interface~(PSI) trees, which are semantically enriched ASTs with type information.
PSI is also tailored towards efficient support for source code changes.
Using the compiler internally, as a library, also saves time during compilation, as it completely skips the compiler startup, which in case for Kotlin compiler is quite slow.

\subsection{Fuzzing oracle}

In any testing process, including fuzzing, we need to address the oracle problem: which behavior of the program under test is and is not a bug?
For compilers, of the most relevance are bugs leading to compiler crashes and miscompilations, therefore, for BBF we decided to focus on them. 
Another flavour of interesting compiler bugs are performance degradations~\cite{kitaura2018random}, but we do not consider them in this work.

Kotlin compiler errors can be divided into 3 main groups: front-end errors, backend errors, and miscompilation errors.
With the former two groups, everything is pretty simple: their oracle is whether the compiler crashes or not.
Detecting miscompilation errors is a little more complicated.

Kotlin supports several different target platform, namely, JVM, JavaScript and Native.
Even more so, platforms have various compilation options, which may significantly change the compilation process and its results.
To find miscompilation errors across all possible configurations we need a unified approach, as supporting a separate oracle for every configuration is close to infeasible.

To check for miscompilations, we decided to implement a lightweight instrumentation of the target program at the level of basic blocks, together with tracking of the program state.
If the traces and/or program states differ between two configurations, it signifies a possible miscompilation.
However, we also need to account for platform-specific false positives, i.e., when the program behavior is different because of the specifics of the platform.
For example, an expression \texttt{1 / 0} crashes the JVM, but returns \texttt{undefined} in JavaScript.
Currently, we have preliminary support of such situations, which in many cases requires manual intervention, but we hope to explore ways of handling them in a fully automatic fashion in the future.

\subsection{Post-processing}

As with most successful fuzzing tools, BBF finds a large number of bug-inducing inputs, and without additional post-processing it would be impossible to handle them manually.
Two most widely used post-processing steps are input reduction~\cite{zeller1999yesterday} and bug deduplication~\cite{chen2013taming}.

Input reduction is implemented in order to make a small and understandable test from a possibly large and complex input leading to a compiler error.
The approach we have implemented, called Reduktor, is described in details in~\cite{stepanov2019reduktor}.
On a very high level, it is based on a hybrid approach which combines several language-agnostic and language-specific techniques.
In practice, it effectively removes over 90\% of the irrelevant information, greatly simplifying its subsequent analysis.
The Reduktor effectiveness is supported by the fact that none of the reported bugs had to be significantly manually reduced.

Bug deduplication is another important step in bug analysis.
Many of the bugs found by fuzzing are duplicates, i.e., different instances of the same original bug.
Currently, this research area is still developing~\cite{chen2013taming}, so we opted to use deduplication algorithms based on crash comparisons.
As we do not have any crashes for miscompilations, we do not currently support their deduplication.
However, we are currently working on advanced deduplication algorithms based on fault localization~\cite{wong2016survey}, which should both improve the quality of regular crash deduplication and also support deduplication of miscompilations.

\section{Evaluation}
\label{sec:evaluation}
\urldef\myurl\url{https://youtrack.jetbrains.com/issues?q=%23found-by-fuzzer}

\begin{table*}[tb]
\center
\begin{tabular}{| c | c | c | c | c |}
\hline

\multirow{1}{*}{\bf } & \multicolumn{1}{c|}{\bf NO SEVERITY} & \multicolumn{1}{c|}{\bf MINOR} & \multicolumn{1}{c|}{\bf NORMAL} & \multicolumn{1}{c|}{\bf MAJOR} \\
\hline
Frontend
& 2 & 0 & 0 & 0  \\
\hline
Backend
& 0 & 1 & 5 & 4 \\
\hline
Miscompilation
& 0 & 0 & 2 & 4 \\
\hline
\end{tabular}
\caption{Posted bug severity from the Kotlin YouTrack}
\label{tab:res-bugs}
\end{table*}

\begin{table*}[tb]
\center
\begin{tabular}{| c | r | r | r | r | r | r | r |}
\hline

\multirow{1}{*}{\bf Results} & \multicolumn{1}{c|}{\bf M} & \multicolumn{1}{c|}{\bf G} & \multicolumn{1}{c|}{\bf EM} & \multicolumn{1}{c|}{\bf SPE} & \multicolumn{1}{c|}{\bf TCE} & \multicolumn{1}{c|}{\bf TCE + EM }  \\
\hline
Correct programs, \%
& 11,9 & 0,05 & 10,7 & 24,2 & 63,4 & 13,4\\
\hline
Interesting bugs, \%
& 25,0 & 0,0 & 20,0 & 100,0 & 100,0 & 16,6\\
\hline
Frontend crashes
& 3 & 212 & 4 & 0 & 0 & 5\\
\hline
Backend crashes
& 9 & 0 & 11 & 2 & 3 & 7\\
\hline
Miscompilations
& 0 & 0 & 0 & 0 & 3 & 0\\
\hline
Duplicates
& 49 & 77 & 38 & 1 & 4 & 14\\
\hline
\end{tabular}
\caption{Evaluation results}
\label{tab:res-time}
\end{table*}

To evaluate our approach we conducted two set of experiments.
In the first experiment, aimed at assessing its practical applicability, we ran our tool for 2 weeks and posted all interesting bugs to the Kotlin YouTrack%
\footnote{\myurl}.
For the second experiment, which is designed to compare our approach against the current state-of-the-art, we implemented several compiler fuzzing techniques and ran them with a time budget of 12 hours, after which we analyzed the statistics of the bugs found.
Here is a list of approaches we chose for this experiment.
\begin{itemize}
    \item (M) Mutation-based fuzzing~\cite{holler2012fuzzing};
    \item (G) Grammar based generation~\cite{purdom1972sentence};
    \item (EM) M + language-specific mutations;
    \item (SPE) Skeletal program enumeration~\cite{zhang2017skeletal};
    \item (TCE) Type-centric enumeration;
    \item (TCE + EM) TCE + language-specific mutations.
\end{itemize}
To further enhance mutation-based approaches, we decided to augment language-agnostic mutations with language-specific ones.
These mutations consist of over 20 text- and tree-based transformation, which are based on our previous experience on Kotlin fuzzing and reduction~\cite{stepanov2019reduktor}.

\subsection{Experimental setup}

For the test bench, we used a machine with Intel Core i7-4790 3.6~GHz processor, 32~GB of memory and Intel 535 SSD storage.
The Kotlin compiler version under test was~1.4.10.
For different compiler configurations, we focused on the following three: default JVM backend, JVM backend in Intermediate Representation~(IR) mode, and JavaScript backend.
IR mode is the new backend implementation, which is currently under active development, and finding its bugs and miscompilations is of much importance to the compiler team.
As programs for initial seed corpus, we selected the Kotlin compiler test suite, consisting of more than 5000 test files.

We discussed some of the implementation details in section~\ref{sec:implementation}, and the fuzzing process follows them.
After generating the input program, we check it for correctness.
If it crashes the compiler, we record the crash and continue.
If it is semantically valid, we instrument the program and execute it, collecting the traces and states.
After the executions are done for all three configurations, their results are compared to each other, with possible miscompilations reported.

\subsection{Interesting bugs found}

For our first experiment, we ran TCE for 2 weeks and found more than 50 unique, \emph{previously not reported} bugs after their post-processing and deduplication, but most of them did not meet the ``interestingness'' criteria from the compiler team~(i.e., not possible in a real-world project or not major enough).
An example of such uninteresting bug is shown in figure~\ref{fig:res-example-3}.

Table~\ref{tab:res-bugs} summarizes 21 \emph{interesting} bugs we found.
All bugs were accepted, 3 of them were actually found to be duplicates.
As seen from the table, 8~(44\%) bugs received major severity, 15~(83\%) received normal or higher severity, meaning over 80\% of the found bugs were deemed important by the compiler team.
Thus, TCE was successful in finding a large number of significant and useful bugs in the current version of the Kotlin compiler in a relatively short time.
We plan to continue our fuzzing efforts in the future, working towards a fully automatic fuzzing pipeline, which can be included in the compiler continuous integration process.
This requires significant improvements to deduplication algorithms and crash/miscompilation oracle construction. 

As an example of interesting bugs found, here are a couple of samples.
Code from figure~\ref{fig:res-example-1} causes a crash in the IR backend, because of a complex expression~(\texttt{5.inv()}) being used as an~\texttt{add} function argument.
TCE can generate such inputs as it is type-aware.
The second code sample in figure~\ref{fig:res-example-2} leads to an IR backend crash due to its complicated behaviour w.r.t. overload resolution.
This bug was triaged as major, actively worked on and has already been fixed.

\subsection{Comparison of TCE with other approaches}

The comparison results of various fuzzing techniques on the Kotlin compiler run for 12 hours are presented in table~\ref{tab:res-time}.
We can highlight the following insights from our comparison.
In quantitative terms, for backend errors mutation-based fuzzing~(M, EM) gives the best results, but most of found bugs are ``uninteresting''.
Nevertheless, about 20-25 percent of bugs are considered ``interesting'', albeit trivial, and the high number of duplicates suggests that these bugs are relatively shallow and easy to find.
As expected, approaches based on mutations have difficulties with finding miscompilation bugs, since the resulting code is usually not very complex.

The grammar-based approach, not surprisingly, are best at finding front-end errors, but managed to generate less than 10~(0.05\%) semantically correct programs.
All 212 bugs found are semantically incorrect and considered uninteresting by the compiler team.
In our opinion, grammar-based approaches are better suited for augmenting other approaches, e.g., generating code snippets for mutation-based or hybrid approaches.

\begin{figure}[tb]
\begin{lstlisting} 
val range1 = 1 downTo 1;
fun testR1xE1() = range1 in 1 downTo 1 
\end{lstlisting}
\caption{Bug found by SPE}
\label{fig:spe-bug-example}
\end{figure}

SPE, in comparison with TCE, gives poorer results.
From our analysis, this happens because the test suite for the Kotlin compiler, used for seed programs, is less suitable for SPE: almost all tests are very simple and contain very few variables, whereas SPE is based around enumerating different variable patterns.
Nevertheless, for some seed programs it managed to find bugs, all of them were interesting~(which proves the usefulness of SPE for real-world testing).
Figure~\ref{fig:spe-bug-example} shows one of SPE bugs.
It was generated by replacing another variable~(deleted by Reduktor afterwards) with~\texttt{range1}.

\begin{figure}[tb]
\begin{lstlisting} 
fun f(): String {
    val p1 = "K"

    open class Kla1(val x: String) {
        constructor() : this("O")
        val p2: String = p1
    }

    class Kla2 : Kla1()
    val p3 = Kla2()

    return p3.x + p3.p2
}

// JVM output: "OK"
// JS  output: "undefined"
\end{lstlisting}
\caption{Miscompilation found by TCE}
\label{fig:TCE-example}
\end{figure}

TCE is the only approach which managed to find actual miscompilation bugs~(one of these bugs is shown in figure~\ref{fig:TCE-example}), and is also best at generating semantically valid code.
However, due to the complexity of Kotlin type system, its implementation in the compiler and other implementation difficulties, still about 35\% of the generated tests were not correct.
Improving this statistic is our primary direction of the future work.

\section{Discussion}
\label{sec:discussion}

\subsection{Generalizability of type-centric enumeration}

When we devised and implemented TCE, we were mainly concerned with its applicability to Kotlin, as it is our main research subject.
However, we believe it to be in the middle ground between language-agnostic and language-specific fuzzing techniques.
The only language-specific components TCE requires are types, which could be provided by either the compiler itself~(in case of a statically typed language) or by a separate type analysis~(if we are working with a dynamically typed language).
In our future work we plan to implement TCE for other programming languages to measure how language-agnostic it is in practice.

\subsection{Seed selection}

Our evaluation results show that approaches such as SPE or TCE are sensitive to the initial program seeds; for example, on Kotlin test suite SPE performs relatively poorly, as the programs contain very few variables, which inhibits the SPE enumeration ability.
Generally, this is true for most mutation-based fuzzing approaches~\cite{yoo2012regression}.

Therefore, it would be interesting to allow for some kind of pre-processing, which can measure if the initial seeds are suitable for the selected fuzzing technique.
Another alternative is to use test suite prioritization~\cite{chen2017learning} or reduction~\cite{chae2011automated} algorithms to better focus the fuzzing efforts.
This requires us to get the test coverage information, which has the benefit of supporting different flavours of coverage-guided fuzzing~\cite{pacheco2007feedback}.

\subsection{Project-level fuzzing}

When working with programs, TCE merges two separate files together: one from the generative phase and one from the mutation phase.
However, real-world projects usually consist of many files with complicated interdependencies.
In principle, TCE could be extended to support merging more than two files together, enumerating over possible slices of project parts, adaptively selecting two sets of files to use for generation and mutation phases.

\section{Related work}
\label{sec:related-work}

Researchers have been applying fuzzing to compilers for quite some time now.
In this section, we review the most relevant approaches and highlight the most important differences with our approach.
For further reading, we redirect you to a more detailed overview of fuzzing in~\cite{chen2020survey}.

Historically, the first approaches to compiler fuzzing were based on grammar-directed test generation, as in~\cite{purdom1972sentence} fuzzing is used for parser testing.
There are a lot of grammar-based works~\cite{duncan1981using, zelenov2003test, havrikov2019systematically}, most of which attempt to enrich the grammar with additional information, to improve the probability of generating a semantically correct program.
However, for complex languages~(such as C++ or Kotlin) even the most advanced grammar-based approaches cannot produce semantically correct inputs with a satisfactory success rate.

Language-specific fuzzers deal with this problem by being language-specific, i.e., incorporating deep knowledge of the target programming language; CSmith~\cite{yang2011finding} is the most well-known tool of this kind.
It is a generator of C programs, with the generation rules hand-written programmatically, so that the generated programs are free of undefined behavior and are syntactically and semantically correct.
But developing such a tool for a new language from scratch is a very long and tedious process.
Besides CSmith, there are generative fuzzers for OpenCL~\cite{lidbury2015many}, JavaScript~\cite{ruderman2007introducing}, and Java~\cite{javaFuzzer}.

Recently, in~\cite{kreutzer2020language} the authors proposed an approach to generate semantically valid code from a language specification written in the LaLa language, describing both syntactic and semantic rules of the target language.
In our future work, we plan to develop a LaLa specification for Kotlin and compare TCE with this approach.

Another option is to use mutation-based fuzzing, for example, LangFuzz~\cite {holler2012fuzzing} or IFuzzer~\cite{veggalam2016ifuzzer}, which processes a number of seed programs and then iteratively mutates them, switching nodes of the same types with each other.
Another example of such an approach is~\cite{chen2016coverage} aimed at finding bugs in JVM implementations.
To do this, the authors implemented a large selection of mutations over~\texttt{.class} files, for example, ``insert / delete a class field'', ``insert / delete a method'' or ``add throwing an exception''.
The mutation phase of TCE is heavily inspired by these approaches and uses the same ideas.

In~\cite{zhang2017skeletal} the authors propose skeletal program enumeration, which we discuss in more detail in section~\ref{sec:spe}.
Our approach is an extension of this idea, on which we improve on by enhancing the enumeration process with type awareness.

A large number of mutation-based compiler fuzzers are based on the idea of equivalence modulo inputs~(EMI)~\cite{le2014compiler}, a variant of metamorphic testing, which checks that behaviour-preserving program mutations are respected by the compiler, i.e., do not change the program behaviour.
Orion~\cite{le2014compiler} mutates parts of a C program not touched during its execution, mutations supported include deletion and addition of such dead code.
Approach~\cite{sun2016finding}, in some sense dual to Orion, is based on adding code to live code regions.
This is done by synthesizing~\texttt{if} expressions with conditions definitely known to be~\texttt{true} or~\texttt{false}.

Of course, there has been a lot of other research in compiler fuzzing.
Several works~\cite{amodio2017neural, cummins2018compiler} apply machine learning to improve the fuzzing performance.
CodeAlchemist~\cite{han2019codealchemist} is based on the idea of semantic-aware assembly of JavaScript programs.
A compiler input is built from existing blocks, annotated with limited semantic information, specifically, which variables are provided by a block and which variables are required by it to be correct.
This approach is ill-suited for languages with complex type systems, as this greatly increases the search space for code block combination algorithm.

\section{Conclusion}
\label{sec:conclusion}

Fuzzing is applicable to most programming languages, but even with the latest advances in fuzzing techniques, generating a non-trivial semantically correct program remains a hard challenge.
In this paper we propose type-centric compiler fuzzing in the form of \emph{type-centric enumeration}~(TCE): an approach inspired by skeletal program enumeration~\cite{zhang2017skeletal}, which extends it to become type-aware, to increase the probability of generating a valid program.
We achieve this by creating a typed skeleton with type placeholder, which are later filled with code fragments of suitable type.

TCE have been implemented for the Kotlin compiler in our fuzzing tool~Backend Bug Finder and extensively evaluated.
For the first part of the evaluation, we fuzzed the compiler using TCE for two weeks and found 18 previously unknown bugs, which were approved as interesting by the compiler team.
For the second part, we compared TCE to other state-of-the-art fuzzing approaches with runs on a 12 hour time budget.
The results show TCE compares favourably to existing approaches when generating semantically correct code, however, to achieve the best overall results one should utilize a combination of fuzzing techniques, to compliment their strengths.

For our future work, there are several research directions we would like to explore.
First, as discussed in section~\ref{sec:related-work}, approach~\cite{kreutzer2020language} based on LaLa language also focuses on generating semantically valid code, and we plan to develop a LaLa specification for Kotlin and do a comparison with~TCE.

We would also like to attempt implementing TCE for other programming languages and their compilers, especially for dynamically typed ones, to get a better understanding of how flexible and performant the approach is across different kinds of languages with different language features.
Besides other languages, we believe TCE could be applied to other fuzzing problems, i.e., it could be used to automatically find compiler performance degradations~\cite{kitaura2018random}.

\bibliographystyle{IEEEtran}
\bibliography{IEEEabrv,bbf}

\balance

\begin{figure*}
\begin{subfigure}[b]{0.45\linewidth}
\begin{lstlisting}[
  language=Kotlin,
  escapechar=|
]
class Kl0 {
    operator fun set(i1: Int, 
                     i2: Int = 1234, 
                     v: String) {
        println(i2)
    }
}

fun main() {
    Kl0()[1] = "OK"
}
\end{lstlisting}
\caption{Miscompilation in default backend~(KT-42064)}
\label{fig:res-1}
\end{subfigure}%
\hfill
\begin{subfigure}[b]{0.45\linewidth}
\begin{lstlisting}[
  language=Kotlin,
  escapechar=|
]
class Kl0 {
    fun f13(): Int = TODO()
}
fun f1(fn: suspend () -> Int): Any 
    = TODO()

fun box() {
    f1(Kl0()::f13)
}
\end{lstlisting}
\caption{IR backend crash~(KT-42021)}
\label{fig:res-2}
\end{subfigure}
\hfill
\begin{subfigure}[b]{0.45\linewidth}
\begin{lstlisting}[
  language=Kotlin,
  escapechar=|
]
class L(val x: String) {
    val xx get() = x
}

fun main() {
    L(L("OK")::xx.get())::xx.get()
}
\end{lstlisting}
\caption{IR backend crash~(KT-42354)}
\label{fig:res-3}
\end{subfigure}
\hfill
\begin{subfigure}[b]{0.45\linewidth}
\begin{lstlisting}[
  language=Kotlin,
  escapechar=|
]
fun main() {
    val a: UByte = 10u
    val b: UByte = 0u
    for (i in a until b) {
        println("a")
    }
}
\end{lstlisting}
\caption{Miscompilation in IR backend~(KT-42186)}
\label{fig:res-4}
\end{subfigure}
\hfill
\begin{subfigure}[b]{0.45\linewidth}
\begin{lstlisting}[
  language=Kotlin,
  escapechar=|
]
class Kl1 : HashSet<String>()
open class Kl2(par0: Any, par1: Any)

fun box1() =
    object : Kl2(
        par1 = "",
        par0 = Kl1().iterator().next()
    ) {}
\end{lstlisting}
\caption{Miscompilation in default backend~(KT-42064)}
\label{fig:res-5}
\end{subfigure}%
\hfill
\begin{subfigure}[b]{0.45\linewidth}
\begin{lstlisting}[
  language=Kotlin,
  escapechar=|
]
fun box() {
    val b: ArrayList<Int?> =
        arrayListOf(1)
    var i = 0
    b[i] == ++i
}
\end{lstlisting}
\caption{IR backend crash~(KT-42251)}
\label{fig:res-6}
\end{subfigure}
\hfill
\caption{Interesting bugs found by TCE}
\label{fig:res}
\end{figure*}

\begin{appendices}

\section{Examples of found bugs}

Figure~\ref{fig:res-1} demonstrates a miscompliation bug in the default JVM backend.
It is caused by the incorrect generation of code handling the default value of the parameter in an operator function.
Bug was given major priority, but remains unassigned as of the writing of this paper.

Figure~\ref{fig:res-2} shows an IR backend crash, caused by~\texttt{suspend} conversion.
It was generated by creating a function reference and replacing the argument of~\texttt{f1} function with it.
Bug was given a normal priority and has been fixed.

Figure~\ref{fig:res-3} shows a miscompilation in IR backend with nested invocations of~\texttt{KProperty} \texttt{get} function.
By allowing for such nested patterns in~TCE, as described in section~\ref{sec:tce-problems}, we managed to find this bug.
Bug was given major priority and later fixed.

Figure~\ref{fig:res-4} demonstrates an miscompilation in IR backend with unsigned integer loop range.
It causes an infinite loop at runtime.
Bug was given major priority and has been fixed.

Figure~\ref{fig:res-5} shows an IR backend crush, caused by interaction between inheritance, named arguments and default values.
It was generated by TCE by replacing one of the default values with a generated expression.
Bug was given normal priority and has been quickly fixed. 

Figure~\ref{fig:res-6} demonstrates an IR backend crash when comparing elements of types~\texttt{Int?} and~\texttt{i}.
Bug was given major priority and has also been fixed.

\end{appendices}

\end{document}